\documentclass[mathleft]{an}
\usepackage{graphicx,bm}
\usepackage{times}
\graphicspath{{./fig/}{./png/}}
\newcommand{\BoldVec}[1]{\mathchoice%
  {\mbox{\boldmath $\displaystyle     #1$}}%
  {\mbox{\boldmath $\textstyle        #1$}}%
  {\mbox{\boldmath $\scriptstyle      #1$}}%
  {\mbox{\boldmath $\scriptscriptstyle#1$}}%
}
\newcommand{\EQ}{\begin{equation}}
\newcommand{\EN}{\end{equation}}
\newcommand{\EQA}{\begin{eqnarray}}
\newcommand{\ENA}{\end{eqnarray}}
\newcommand{\eq}[1]{(\ref{#1})}

\newcommand{\Eq}[1]{Eq.~(\ref{#1})}
\newcommand{\Eqs}[2]{Eqs~(\ref{#1}) and~(\ref{#2})}

\newcommand{\Fig}[1]{Fig.~\ref{#1}}
\newcommand{\FFig}[1]{Figure~\ref{#1}}
\newcommand{\Tab}[1]{Table~\ref{#1}}
\newcommand{\Figs}[2]{Figs.~\ref{#1} and \ref{#2}}

\newcommand{\bra}[1]{\langle #1\rangle}

\newcommand{\xx}{\BoldVec{x}{}}

\newcommand{\uu}{\BoldVec{u} {}}

\newcommand{\qq}{\BoldVec{q} {}}

\newcommand{\eee}{\BoldVec{e} {}}

\newcommand{\ff}{\BoldVec{f} {}}

\newcommand{\GG}{\BoldVec{G} {}}
\newcommand{\kk}{\BoldVec{k} {}}

\newcommand{\nab}{\BoldVec{\nabla} {}}

\newcommand{\oo}{\BoldVec{\omega} {}}

\newcommand{\SSSS}{\bm{\mathsf{S}}}

\newcommand{\ii}{{\rm i}}

\newcommand{\dd}{{\rm d} {}}
\newcommand{\const}{{\rm const}  {}}

\def\Ma{\mbox{\rm Ma}}

\def\Rey{\mbox{\rm Re}}

\def\cs{c_{\rm s}}
\def\kd{k_{\rm d}}
\def\kf{k_{\rm f}}
\def\kT{k_{\rm Tay}}
\def\ke{k_{\rm eff}}
\def\tkd{\tilde{k}_{\rm d}}
\def\tkf{\tilde{k}_{\rm f}}
\def\tkT{\tilde{k}_{\rm Tay}}
\def\tke{\tilde{k}_{\rm eff}}
\def\urms{u_{\rm rms}}

\def\half{{\textstyle{1\over2}}}

\def\onethird{{\textstyle{1\over3}}}

\newcommand{\yjgr}[3]{: #1, {JGR} {#2}, #3}

\newcommand{\yapj}[3]{: #1, {ApJ} {#2}, #3}

\newcommand{\yana}[3]{: #1, {A\&A} {#2}, #3}

\newcommand{\ygafd}[3]{: #1, {GApFD} {#2}, #3}
\newcommand{\yjfm}[3]{: #1, {JFM} {#2}, #3}
\newcommand{\ypf}[3]{: #1, {PhFl} {#2}, #3}

\newcommand{\yprl}[3]{: #1, {PhRvL} {#2}, #3}
\newcommand{\ypre}[3]{: #1, {PhRvE} {#2}, #3}

\newcommand{\ymn}[3]{: #1, {MNRAS} {#2}, #3}

\newcommand{\yjour}[4]{: #1, {#2} {#3}, #4}

\newcommand{\ybook}[3]{: #1, {\it #2} (#3)}

\begin{document}

\Pagespan{195}{}
\Yearpublication{2012}%
\Yearsubmission{2012}%
\Month{3}%
\Volume{333}%
\Issue{3}%
\DOI{10.1002/asna.201211654}

\title{Kinetic helicity decay in linearly forced turbulence}

\author{A. Brandenburg\inst{1,2}\fnmsep\thanks{Corresponding author:
\email{brandenb@nordita.org}\newline}
\and A. Petrosyan\inst{3,4}}

\titlerunning{Kinetic helicity decay}

\authorrunning{A. Brandenburg \& A. Petrosyan}

\institute{
NORDITA, AlbaNova University Center,
Roslagstullsbacken 23, SE-10691 Stockholm, Sweden
\and
Department of Astronomy, AlbaNova University Center,
Stockholm University, SE 10691 Stockholm, Sweden
\and
Space Research Institute of the Russian Academy of Sciences
Profsoyuznaya 84/32, Moscow 117997, Russia
\and
Moscow Institute of Physics and Technology, State University,
Institutsky lane 9, Dolgoprudny 141700, Russia
}

\received{2012 Feb 24} \accepted{2012 Mar 9} \publonline{2012 Apr 5}


\abstract{
The decay of kinetic helicity is studied in numerical models of forced
turbulence using either an externally imposed forcing function as an
inhomogeneous term in the equations or, alternatively, a term linear
in the velocity giving rise to a linear instability.
The externally imposed forcing function injects energy at the largest
scales, giving rise to a turbulent inertial range with nearly constant
energy flux while for linearly forced turbulence the spectral energy
is maximum near the dissipation wavenumber.
Kinetic helicity is injected once a statistically steady state is reached,
but it is found to decay on a turbulent time scale regardless of the
nature of the forcing and the value of the Reynolds number.
}

\maketitle

\section{Introduction}
\label{Introduction}

The physical properties of astrophysical turbulence are often studied
by solving the hydrodynamic equations in a periodic domain with an
assumed forcing function.
In particular,
isotropic homogeneous turbulence is often studied as a proxy of turbulence
in more complicated situations, where specific concepts and general
aspects are harder to isolate.
An important concept is that of the forward cascade of kinetic energy
to smaller scales.
This leads to a $k^{-5/3}$ energy spectrum, where $k$ is the wavenumber.
Many flows of geophysical and astrophysical relevance are subject to
rotation and stratification and can therefore attain kinetic helicity.
Closure calculations (\cite{AL77}), direct numerical simulations
(\cite{BO97}; \cite{BS05a}), as well as shell model calculations
(\cite{Chkhetiani}; \cite{DG01},\cite{DG01b}; \cite{Stepanov}) show
an approximate $k^{-5/3}$ scaling for the kinetic helicity, suggesting
that kinetic helicity too is subject to a forward cascade toward smaller
length scales.
On the other hand, kinetic helicity is conserved by the quadratic
interactions and might therefore play an important role in the inviscid limit.
Although this is also the case in ideal magnetohydrodynamics (MHD),
where magnetic helicity is also conserved by the quadratic interactions,
there is a significant difference.
In the ideal case of small magnetic diffusivity, magnetic helicity can
only evolve on resistive time scales.
This can have profound effects on the saturation behavior of large-scale
dynamos (\cite{B01}).

As mentioned above, in a polytropic flow,
the quadratic interactions conserve kinetic helicity,
$\bra{\oo\cdot\uu}$, where $\oo=\nab\times\uu$ is the vorticity,
$\uu$ the velocity, and angular brackets denote volume averaging
over a closed or periodic domain.
This conservation property becomes evident when writing the
Navier-Stokes equation in the form
\EQ
{\partial\uu\over\partial t}=\uu\times\oo-\nab P+\ff
+\nu[\onethird\nab(\nab\cdot\uu)-\nab\times\nab\times\uu+\GG],
\EN
where $P=\half\uu^2+h$ is the sum of specific turbulent pressure,
$\uu^2/2$, and specific enthalpy, $h=\cs^2\ln\rho$.
Furthermore, $\rho=\const$ is the density, and $\ff$ is a forcing function.
In the absence of forcing, $\ff=\bm{0}$, kinetic helicity is just
subject to viscous decay, because the nonlinear term $\uu\times\oo$
is perpendicular to $\oo$, i.e., we have
\EQ
{\dd\over\dd t}\bra{\oo\cdot\uu}=-2\nu\bra{\qq\cdot\oo},
\EN
where $\qq=\nab\times\oo$ is the curl of the vorticity.
In the ideal case, $\nu=0$, we have $\bra{\oo\cdot\uu}=\const$.
However, in fluid dynamics the ideal case is hardly representative
of the limit of large Reynolds numbers, where $\nu\to0$.
Indeed, for a self-similar decay of kinetic energy, the wavenumber
of the energy-carrying eddies, $\kf$, decreases with time such that
$\nu\kf^2t\approx\const$.
This implies that the rate of energy decay,
$\nu\bra{\oo^2}$, is essentially independent of $\nu$
and hence independent of the Reynolds number.
Given that $\bra{\uu^2}$ is related to the kinetic energy,
which is also independent of the Reynolds number, we expect
that the ratio
\EQ
\bra{\oo^2}/\bra{\uu^2}\equiv\kT^2,
\label{kT}
\EN
which is related to the Taylor micro-scale wavenumber $\kT$,
should be proportional to $\Rey$, and therefore
\EQ
\kT\sim\Rey^{1/2}.
\label{kscaling}
\EN
However, for helical flows the rate of kinetic helicity dissipation is
proportional to $\nu\bra{\qq\cdot\oo}$.
Thus, if we define
\EQ
\bra{\qq\cdot\oo}/\bra{\oo\cdot\uu}\equiv\ke^2,
\label{ke}
\EN
we see that kinetic helicity dissipation is related to kinetic
energy by altogether 3 wavenumber factors.
If all these factors scale like in \Eq{kscaling}, we
may expect the rate of kinetic helicity dissipation
to diverge with decreasing $\nu$ like $\nu^{-1/2}$.
This is in stark contrast to the related case of magneto-hydrodynamic
turbulence where, following similar reasoning, the magnetic helicity
dissipation converges to zero like $\eta^{1/2}$ as the magnetic
dissipation $\eta$ goes to zero (\cite{BS05}); see also the appendix
of \cite{BK07} for a clear exposition of these differences.

A problem with the simple argument above is that in cases of practical
relevance the forcing function $\ff$ usually breaks kinetic
helicity conservation.
This is particularly evident for the so-called linear forcing model
of \cite{Lun03}, where
\EQ
\ff=A\uu
\label{LinForcing}
\EN
is a positive multiple of the velocity vector.
In that case we have
\EQ
{\dd\over\dd t}\bra{\oo\cdot\uu}=2A\bra{\oo\cdot\uu}-2\nu\bra{\qq\cdot\oo},
\label{ExpIncrease}
\EN
so that $\bra{\oo\cdot\uu}$ could even exhibit exponential growth.
An aim of this paper is thus to investigate to what degree kinetic helicity is
conserved in forced turbulence using both the linear forcing model and compare
it with the more traditional stochastic forcing in a narrow wavenumber band.
Another motivation is the fact that, by analogy, magnetic helicity turned
out to be of crucial importance in understanding the saturation properties
of helically forced dynamos in periodic domains (see, e.g., \cite{B01}).
We study the kinetic helicity evolution by monitoring the response
to adding a large-scale helical component to the flow for both types
of forcing.

\section{Method}

In the following we present results obtained by solving the compressible
hydrodynamic equations with an imposed random forcing term and an
isothermal equation of state, so that the pressure $p$ is related to
$\rho$ via $p=\rho\cs^2$, where $\cs$ is the isothermal sound speed.
We thus deviate from the conceptually simpler incompressible case.
The reason for doing this is that we are mainly interested in astrophysical
applications where the gas is compressible.
Furthermore, we expect that for small Mach numbers, compressible and
incompressible cases become nearly identical.
However, this is only partially true, as the zero-Mach number limit may
not be uniform and may not commute with the small-scale, long-time or
zero-viscosity limits.

In the following we use an isothermal equation of state for a
monatomic gas for which the bulk viscosity vanishes, so the
hydrodynamic equations for $\rho$ and $\uu$ take the form
\EQ
{\partial\rho\over\partial t}=-\nab\cdot\rho\uu,
\EN
\EQ
{\partial\uu\over\partial t}=-\uu\cdot\nab\uu-\cs^2\nab\ln\rho
+\ff+\rho^{-1}\nab\cdot2\rho\nu\SSSS,
\label{dUdt}
\EN
where ${\sf S}_{ij}=\half(u_{i,j}+u_{j,i})-\onethird\delta_{ij}\nab\cdot\uu$
is the traceless rate of strain tensor, $\nu$ is the kinematic viscosity,
and $\ff$ is a for\-cing function
that is either given by the linear forcing model using \Eq{LinForcing}
with a positive constant $A$,
or, alternatively, it is given by a random forcing function consisting of
plane transversal waves with random wavevectors $\kk$ such that $|\kk|$
lies in a band around a given forcing wavenumber $k_{\rm f}$.
The vector $\kk$ changes randomly from one timestep to the next,
so $\ff$ is $\delta$ correlated in time.
The forcing amplitude is chosen such that the Mach number, $\Ma=\urms/\cs$,
is about 0.1.
Here, $\urms=\bra{\uu^2}^{1/2}$ is the root-mean-square (rms) velocity.
We use triply-periodic boundary conditions in a Cartesian domain of size $L^3$.
The smallest wavenumber that fits into the computational domain
is $k_1=2\pi/L$.

For the linear forcing model we choose for the amplitude $A/\cs k_1=0.02$,
while for the random forcing function, $\ff$ is of the form
\EQ
\ff(\xx,t)=\Rey\{N\ff_{\kk(t)}\exp[\ii\kk(t)\cdot\xx+\ii\phi(t)]\},
\EN
$\xx$ is the position vector and
\EQ
\ff_{\kk}=\left(\kk\times\eee\right)/\sqrt{\kk^2-(\kk\cdot\eee)^2},
\label{nohel_forcing}
\EN
where $\eee$ is an arbitrary unit vector not aligned with $\kk$;
note that $|\ff_{\kk}|^2=1$.
The wave vector $\kk(t)$ and the random phase
$-\pi<\phi(t)\le\pi$ change at every time step.
For the time-integrated forcing function to be independent
of the length of the time step $\delta t$, the normalization factor $N$
has to be proportional to $\delta t^{-1/2}$.
On dimensional grounds it is chosen to be
$N=f_0 c_{\rm s}(\kf c_{\rm s}/\delta t)^{1/2}$, where $f_0$ is a
nondimensional forcing amplitude, which is chosen to be $f_0=0.02$.
For the monochromatically forced simulations we choose $\kf=1.5\,k_1$.
For the linear forcing module, on the other hand, we compute $\kf$
is the integral scale from the resulting kinetic energy spectrum
(see below).

Our simulations are characterized by the value of the Reynolds number,
\EQ
\Rey=\urms/\nu\kf.
\EN
where the wavenumber $\kf$ is either the forcing wavenumber in the
case of monochromatic random forcing, or it is evaluated as the wavenumber
of the energy-carrying eddies,
\EQ
\kf^{-1}=\int k^{-1} E(k)\,\dd k\left/\int E(k)\,\dd k\right.,
\label{kf}
\EN
with $E(k)$ being the kinetic energy spectrum,
which is here normalized such that
\EQ
\int_0^\infty E(k)\,\dd k=\half\bra{\uu^2}.
\EN
It turns out the in the former case of monochromatic random forcing,
the definition \eq{kf} agrees well with the a priori chosen forcing wavenumber.
We plot energy spectra as a function of $k$ in units of the dissipation
wavenumber,
\EQ
k_{\rm d}=(\epsilon/\nu^3)^{1/4}.
\EN
Here, $\epsilon=\nu\bra{\oo^2}$ is the rate of energy dissipation.
We also analyze spectra of kinetic helicity, $F(k)$, which are normalized
such that
\EQ
\int_0^\infty F(k)\,\dd k=\bra{\oo\cdot\uu}.
\EN
We note that the kinetic helicity spectrum can have either sign and its
modulus obeys the well-known realizability condition, $|F(k)|\leq2kE(k)$
(\cite{Mof69}).

In the statistically steady state, the kinetic helicity fluctuates
around zero.
We consider the evolution of kinetic helicity after adding an instantaneous
finite perturbation in the form of a Beltrami field,
\EQ
\uu\to\uu+\epsilon\cs(\sin kz, \cos kz, 0),
\label{Beltrami}
\EN
where $\epsilon$ is a nondimensional input parameter.
In all cases we choose $k=k_1$, i.e., we perturb the system with
a wave on the scale of the domain.
We express time in units of turnover times, so that $t\urms\kf$
is a nondimensional time.

The simulations have been carried out using the {\sc Pencil
Code} \footnote{http://pencil-code.googlecode.com/} which
is a high-order finite-difference code (sixth order in space and third
order in time) for solving the compressible hydrodynamic equations.

\begin{figure}[t!]\begin{center}
\includegraphics[width=\columnwidth]{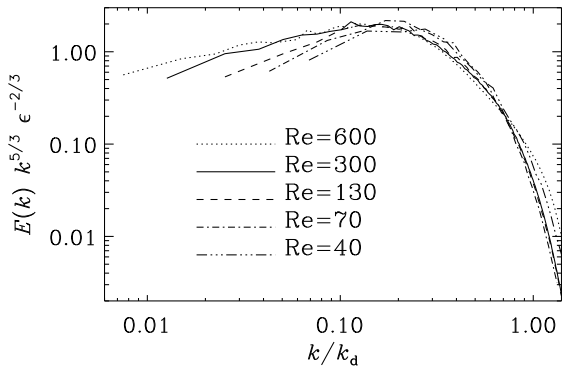}
\end{center}\caption[]{
Kinetic energy spectra compensated with $k^{5/3}\epsilon^{-2/3}$
for a range of different Reynolds numbers and numerical resolutions
up to $512^3$ mesh points.
}\label{pspec_comp}\end{figure}

\section{Results}

\subsection{Linear forcing model}

In \Fig{pspec_comp} we show compensated spectra of
kinetic energy for different values of the Reynolds number.
The spectra collapse onto each other in the dissipation
range if the wavenumber is scaled with $k_{\rm d}$.
With increasing Reynolds number the spectra begin to sketch out
a  continuation toward smaller wavenumber $k$.
Note that in the linear forcing model, energy is being injected at
all length scales and not just at the largest scale of the domain.
This is a property that may also be responsible for the fact
that the effective driving scale tends to be smaller than in
otherwise equivalent monochromatically forced simulations
(\cite{RM05}).

\begin{figure}[t!]\begin{center}
\includegraphics[width=\columnwidth]{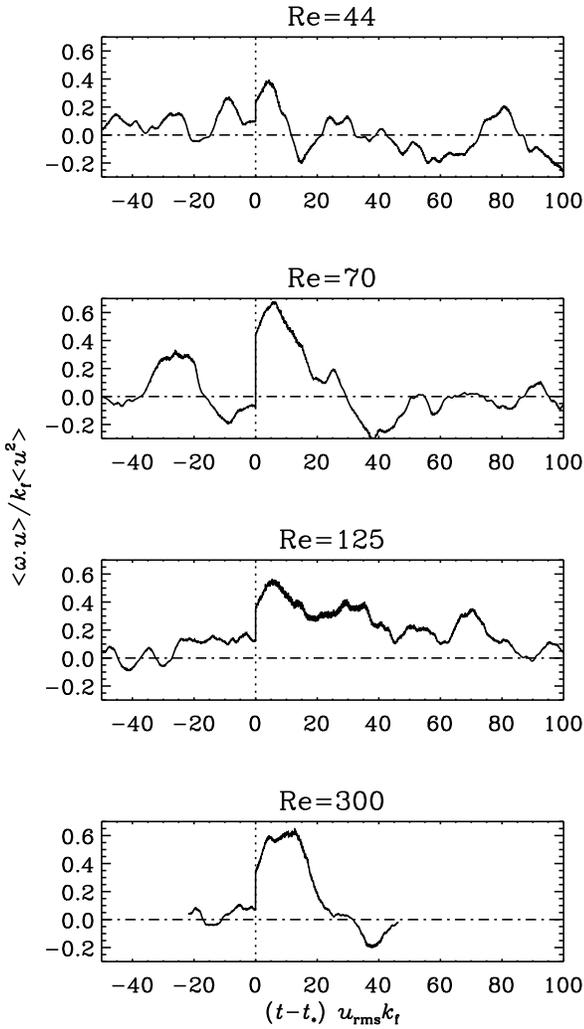}
\end{center}\caption[]{
Evolution of the normalized kinetic helicity for different values
of Re after adding a Beltrami field perturbation at $t=t_*$.
}\label{prelhel_comp}\end{figure}

In the statistically steady state, the kinetic helicity fluctuates
around zero.
However, after adding a finite perturbation in the form of \Eq{Beltrami},
the helicity begins to grow for about 5 turnover times,
$\delta t\urms\kf\approx5$.
After that time there is a systematic decline of the kinetic helicity.
By inspecting the results for different values of $\Rey$, we see that the
time it takes for the kinetic helicity to relax to previous levels
becomes longer as the Reynolds number is increased from 44 to 125;
see \Fig{prelhel_comp}.
However, for $\Rey=300$, which is our largest Reynolds number for which
we have performed experiments with added Beltrami fields, the decline
of helicity has happened on a similar time scale as for $\Rey=70$.
This suggests that there may not be a systematic Reynolds number dependence
of kinetic helicity decay.

\begin{figure}[t!]\begin{center}
\includegraphics[width=\columnwidth]{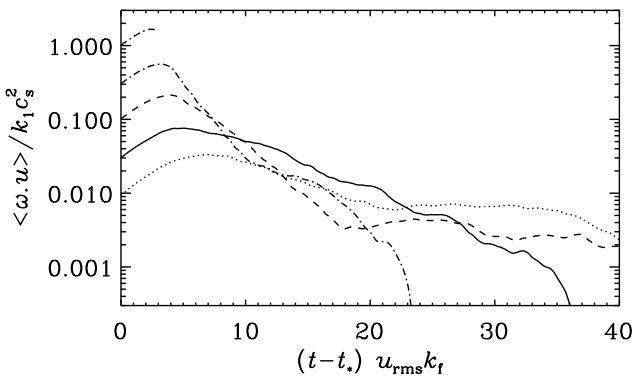}
\end{center}\caption[]{
Evolution of the kinetic helicity for different values of $\epsilon$
after adding the Beltrami field perturbation at $t=t_*$.
}\label{prelhel_comp2}\end{figure}

It should be noted that during the first 2--3 eddy turnover times
after adding the Beltrami field perturbation, the kinetic helicity
shows an exponential increase; see also \Fig{prelhel_comp2}.
This is connected with the fact that in the absence of any other
effective damping, \Eq{ExpIncrease} would imply an exponential growth
of $\bra{\oo\cdot\uu}\propto\exp{\lambda t}$ with growth rate
\EQ
\lambda=2(A-\nu k_{\rm eff}^2),
\label{keff}
\EN
where $k_{\rm eff}$ quantifies the involvement of high wavenumbers in
the expression for the kinetic helicity dissipation; see \Eq{ke}.

One would expect strong perturbations to survive this exponential growth
for longer, but this is not the case, as is demonstrated in \Fig{prelhel_comp2}
where we show the evolution of the kinetic helicity for different values
of $\epsilon$ after adding the Beltrami field perturbation.
The reason for the subsequent decay of magnetic helicity
lies in the fact that $k_{\rm eff}$ scales like $\ke\propto\Rey^{1/2}$,
so the rate of kinetic helicity dissipation remains significant even
in the limit $\Rey\to\infty$, i.e.\ $\nu\to0$.
This is quantified in terms of $k_{\rm eff}$, whose scaling will
with Reynolds number will be considered below.

\begin{figure}[t!]\begin{center}
\includegraphics[width=\columnwidth]{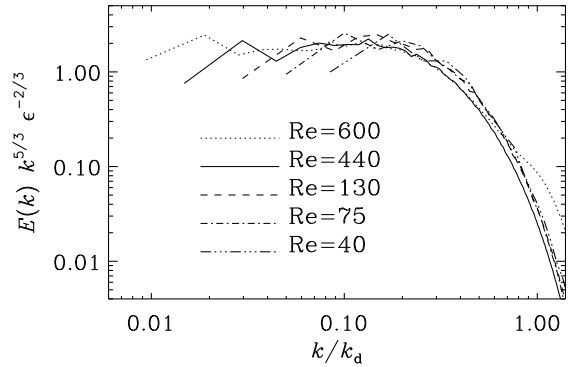}
\end{center}\caption[]{
Kinetic energy spectra compensated with $k^{5/3}\epsilon^{-2/3}$
for a range of different Reynolds numbers and numerical resolutions
using monochromatic forcing.
}\label{pspec_comp_standard}\end{figure}

\subsection{Monochromatic random forcing}

In order to see whether the results presented above are a special
consequence of the linear forcing model, we now perform simulations
using the more traditional monochromatic forcing in a narrow wavenumber
interval.
In \Fig{pspec_comp_standard} we show energy spectra for different Reynolds numbers.
The results suggest that the decline in spectral power toward
the smallest wavenumbers seen in \Fig{pspec_comp} for the linear forcing
model is now absent.
In other words, while in \Fig{pspec_comp_standard} we clearly see that
the compensated energy spectrum is flat, this is not the case in
\Fig{pspec_comp} for the linear forcing model.
However, there is still an uprise near $0.1\,\kd$ that one may generally
associate with the bottleneck effect (\cite{Fal94}; \cite{Dob03}).

\begin{figure}[t!]\begin{center}
\includegraphics[width=\columnwidth]{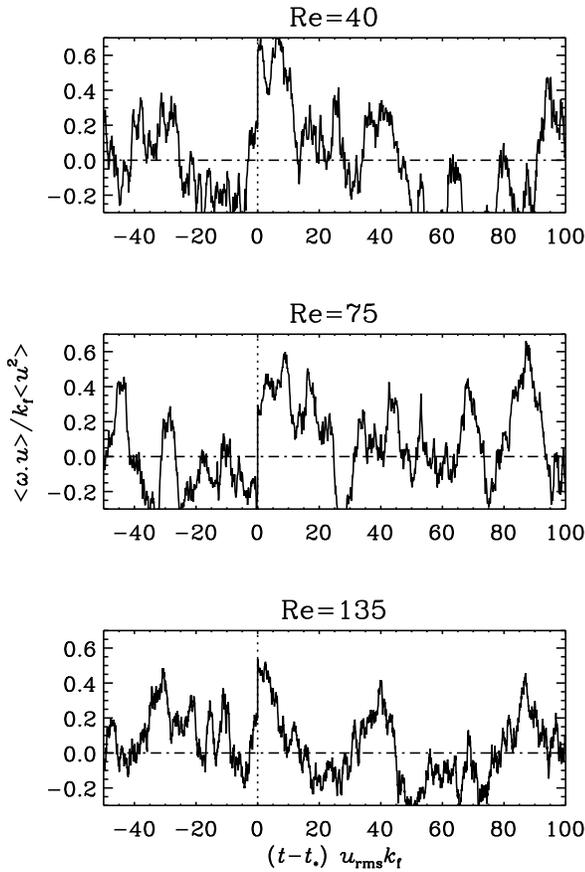}
\end{center}\caption[]{
Evolution of the normalized kinetic helicity
using the monochromatic forcing scheme for different values
of Re after adding a Beltrami field perturbation at $t=t_*$.
}\label{prelhel_comp_standard}\end{figure}

Next, we study the effect of adding a helicity perturbation
also in this case.
\FFig{prelhel_comp_standard} gives time series for three values of $\Rey$.
There is no evidence for a prolonged relaxation to zero.
The reason for this could be that a helical wave cannot interact with
itself; see a corresponding discussion following Eq.~(11) of \cite{Kra73}.
This also suggests that kinetic helicity conservation in the inviscid case,
$\nu=0$, is of no relevance to the inviscid {\it limit}, $\nu\to0$,
in which case the kinetic helicity dissipation diverges.
This behavior was less clear in the previous case with the
linear forcing model.
This might either be a matter of coincidence, but it could also be a
consequence of the linear forcing model which has exponentially
growing solutions.

The time series in \Fig{prelhel_comp_standard}
reveals another interesting aspect in comparison to
\Fig{prelhel_comp} for the linear forcing model in that the level
of fluctuations of $\bra{\oo\cdot\uu}$ is generally larger for the
monochromatic forcing function than for the linear forcing model.
Furthermore, the time series show much stronger short-term fluctuations
while for the linear forcing model the time traces of $\bra{\oo\cdot\uu}$
are smoother.

\begin{figure}[t!]\begin{center}
\includegraphics[width=\columnwidth]{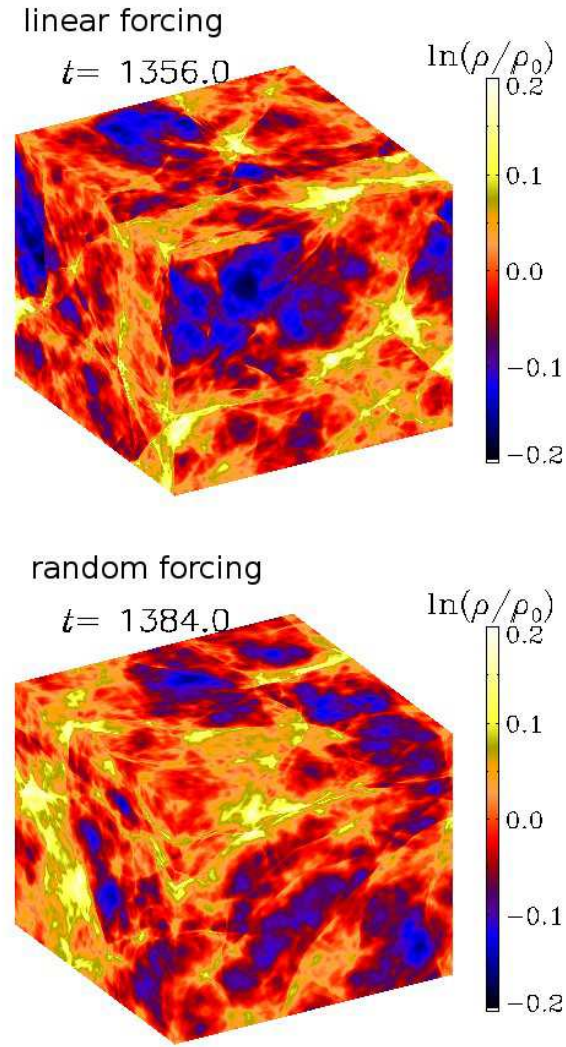}
\end{center}\caption[]{
(online colour at: www.an-journal.org)
Visualization of the logarithmic density (proportional to logarithmic
pressure fluctuations) on the periphery of the domain
for linear and random forcings.
In both cases we have $\Rey\approx600$ at a resolution of $512^3$ mesh points.
}\label{lin_std2}\end{figure}

In \Tab{Ttimescale} we summarize integral and dissipation wavenumbers
as well as the normalized energy fluxes for both linear and monochromatic
random forcings.
In wind tunnel turbulence one usually expresses the energy flux
in units of a quantity $C_\epsilon=u'^3/L$, where $u'=\urms/\sqrt{3}$
is the one-dimensional rms velocity and $L=3\pi/4\kf$ is the customary
definition of the integral scale.
The standard result of $C_\epsilon\approx0.5$ (\cite{pearson}) corresponds
then to $\epsilon\approx0.04\kf\urms^3$.
Comparing the normalized energy fluxes for linear and monochromatic
random forcings we see hardly any differences.
This suggests that the nature of the forcing in hydrodynamic turbulence
might not be of great qualitative importance, although it is still
possible that the bottleneck effect (\cite{Fal94}; \cite{Dob03}) near the
dissipative scale might be connected with the nature of the forcing at
large scales (\cite{Davidson}).

\begin{table*}[t!]\caption{
Summary of the normalized characteristic wavenumbers
$\tkf=\kf/k_1$, $\tkd=\kd/k_1$, $\kT=\kT/k_1$,
and $\tke=\ke/k_1$, for linear and monochromatic forcings.
The numerical resolution is given in the second column.
}\vspace{12pt}\centerline{\begin{tabular}{rr|ccccc|ccccc}
\hline
 &  & \multicolumn{5}{c|}{linear forcing} & \multicolumn{5}{c}{monochromatic forcing} \\
 Re & Res. &$\tkf$&$\tkd$&$\tkT$&$\tke$&$\epsilon/\kf\urms^3$
    &$\tkf$&$\tkd$&$\tkT$&$\tke$&$\epsilon/\kf\urms^3$\\
\hline
 40 &$ 64^3$& 1.6 &   15 &  3 & 3 & 0.080 & 1.6 &  12 &  2 &  3 & 0.072 \\
 70 &$128^3$& 1.9 &   23 &  4 & 5 & 0.069 & 1.9 &  20 &  3 &  4 & 0.052 \\
130 &$128^3$& 2.0 &   40 &  6 & 6 & 0.073 & 1.9 &  34 &  4 &  7 & 0.045 \\
300 &$256^3$& 2.2 &   79 &  9 & 7 & 0.067 & 2.8 &  68 &  7 &  8 & 0.028 \\
600 &$512^3$& 2.2 &  133 & 13 &10 & 0.069 & 1.2 & 130 &  8 &  3 & 0.064
\label{Ttimescale}\end{tabular}}\end{table*}

\begin{figure}[t!]\begin{center}
\includegraphics[width=\columnwidth]{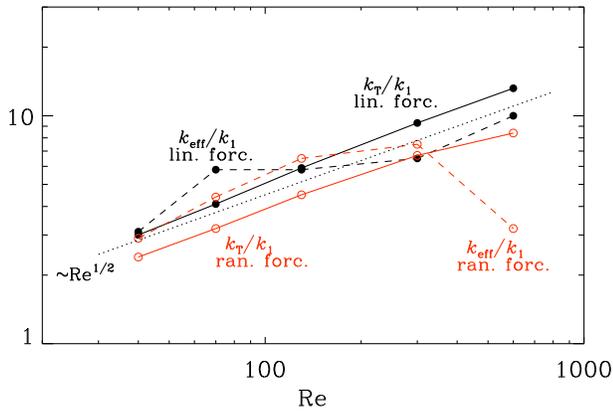}
\end{center}\caption[]{
(online colour at: www.an-journal.org)
Scalings of $\kT/k_1$ (solid lines) and $\ke/k_1$ (dashed lines) with
$\Rey$ for linear (filled symbols) and monochromatic (open symbols and
red lines) forcings.
The dotted line indicates $\Rey^{1/2}$ scaling.
}\label{ptaylor}\end{figure}

In \Tab{Ttimescale} we also list the values of $\kT$ and $\ke$,
as defined in \Eqs{kT}{ke}.
Both wavenumbers are clearly proportional to $\Rey^{1/2}$, as can be seen
from \Fig{ptaylor}, where we plot the Reynolds number dependence of the
ratios $\kT/k_1$ and $\ke/k_1$ for linear and monochromatic forcings.

\begin{figure}[t!]\begin{center}
\includegraphics[width=\columnwidth]{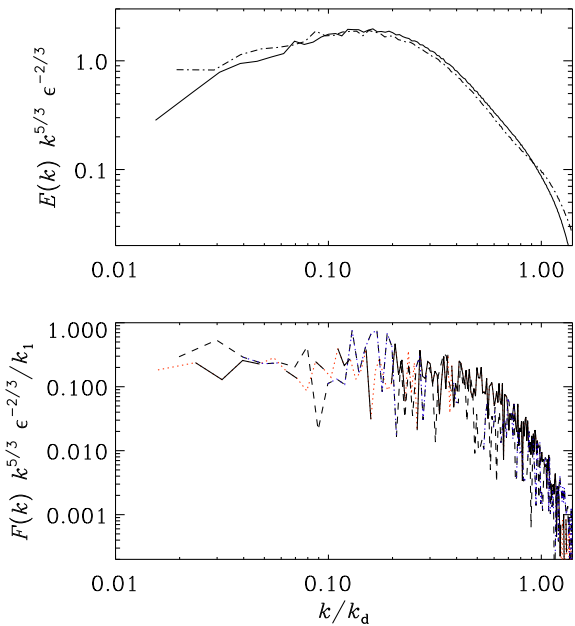}
\end{center}\caption[]{
Comparison of compensated kinetic energy and helicity
spectra for the linear forcing model (solid line) and
the monochromatic random forcing function (dashed).
In those patches where the kinetic helicity is negative,
the modulus of the value is plotted as a dotted line
(red for the linear forcing model and blue for the
monochromatic random forcing function).
}\label{pspec_comp_hel}\end{figure}

To compare the linearly and monochromatically forced turbulence
simulations in real space, we present in \Fig{lin_std2} visualizations
of the logarithmic density on the periphery of the computational domain.
The logarithmic density is a convenient scalar quantity characterizing
the pressure fluctuations resulting from the Reynolds stress.
These visualizations look rather similar in the two cases,
suggesting that the slight difference seen in the power spectra in
\Figs{pspec_comp}{pspec_comp_standard} is hard to discern in such images.
In fact, upon closer comparison of just the cases with the highest
resolution we see that both kinetic energy and kinetic helicity spectra
agree almost perfectly (\Fig{pspec_comp_hel}), except that in the linear
forcing model the kinetic energy spectrum has dropped significantly
at the smallest wavenumber, which is not the case for monochromatic
random forcing.

\section{Conclusions}

The linear forcing model is based on the driving of turbulence by a
linear instability instead of a forcing function that is independent of
the flow.
This type of forcing might be more physical, because it does not change
abruptly and depends on the local flow properties.
Nevertheless, both types of forcing are Galilean invariant.
This would change if the flow-independent monochromatic
forcing were no longer $\delta$ correlated in time.
A disadvantage of linearly forced turbulence is that energy injection
occurs at all wavenumbers.
Our simulations show that the energy spectrum is perhaps slightly shallower
than with the flow-independent monochromatic forcing function.
Part of this is explained by the bottleneck effect (\cite{Fal94}; \cite{Dob03}),
and that the energy spectra compensated by $k^{5/3}$ show a stronger
uprise toward the dissipation wavenumber.
It is unclear whether this result would persist at larger resolution.

The linear forcing model has the interesting property of amplifying
not only kinetic energy, but also kinetic helicity.
Indeed, at early times, just after having injected kinetic helicity into
the system, the coherent helical part continues to increase exponentially,
but the flow soon breaks up into smaller eddies, giving rise to enhanced
effective dissipation.
For practical applications, it should be noted that the linear forcing
model has the disadvantage that the initial exponential growth of kinetic
energy and kinetic helicity might prevail for too long.
This will be the case when the {\it initial} random perturbations are
too weak to perturb the flow sufficiently.
In that case, the kinetic energy would quickly increase to large values
without producing three-dimensional turbulence.

With regards to geophysical and astrophysical applications we can say
that in a turbulent system, kinetic helicity is no longer a conserved
quantity, even though it would be if $\nu=0$ were strictly true.
The latter requirement is of course not really possible in a turbulent
system, because kinetic energy would then accumulate at the smallest
possible scale resolved within the hydrodynamics framework and kinetic
energy would not be able to decay, which is unphysical.
While this should not be surprising, it is important to remember that this
is quite different in the case of magnetohydrodynamics, where magnetic
helicity dissipation really does go to zero in a turbulent system --
even for finite (but small) values of the magnetic diffusivity.
At the same time, magnetic energy dissipation does stay finite
and is able to accomplish magnetic reconnection on the smallest
resolved scales of the turbulent cascade (\cite{GN96}; \cite{LV99}).

This paper has also shown that, regardless of the nature of the forcing,
there are fairly strong helicity fluctuations.
They appear to be coherent over many turbulent eddy timescales.
One may wonder how generic such fluctuations are and if such
fluctuations could be relevant for say the incoherent dynamo effects
that have been investigated by several authors in recent years
(\cite{Vishniac}; \cite{Sur}; \cite{Heinemann}; \cite{Mitra}; \cite{Proctor}).
It will therefore be interesting the associate the kinetic helicity
fluctuations with those of $\alpha$, which have already been determined in
simulations of turbulent shear flows (\cite{BRRK}).

\acknowledgements
We thank two anonymous referees for providing useful comments that
have helped improving the presentation of the paper.
We acknowledge the allocation of computing resources provided by the
Swedish National Allocations Committee at the Center for
Parallel Computers at the Royal Institute of Technology in
Stockholm and the National Supercomputer Centers in Link\"oping
as well as the Norwegian National Allocations Committee at the
Bergen Center for Computational Science.
This work was supported in part by
the European Research Council under the AstroDyn Research Project No.\ 227952
and the Swedish Research Council Grant No.\ 621-2007-4064.


\end{document}